\documentclass[12pt]{elsart}

\usepackage{amsmath}
\usepackage{amsfonts}
\usepackage{graphicx}
\usepackage{amssymb}
\usepackage{ulem}

\begin{document}

\begin{frontmatter}

\title{Study of a quantum scattering process \\ by means of entropic measures}

\author[rlr]{Ricardo L\'{o}pez-Ruiz}
\ead{rilopez@unizar.es} and
\author[jsr]{Jaime Sa\~{n}udo}
\ead{jsr@unex.es}

\address[rlr]{
DIIS and BIFI, Facultad de Ciencias, \\
Universidad de Zaragoza, E-50009 Zaragoza, Spain}

\address[jsr]{
Departamento de F\'isica, Facultad de Ciencias, \\
Universidad de Extremadura, E-06071 Badajoz, Spain, \\
and BIFI, Universidad de Zaragoza, E-50009 Zaragoza, Spain}


\begin{abstract}
In this work, a scattering process of quantum particles through a potential barrier
is considered. The statistical complexity and the Fisher-Shannon information are calculated
for this problem. The behaviour of these entropy-information measures as a function of the energy
of the incident particles is compared with the behaviour of a physical magnitude, the reflection 
coefficient in the barrier. We find that these statistical magnitudes present their minimum values
in the same situations in which the reflection coefficient is null. These are the situations 
where the total transmission through the barrier is achieved, {\it the transparency 
points}, a typical phenomenon due to the quantum nature of the system.
\end{abstract}


\end{frontmatter}

\maketitle

The study of the crossing of potential barriers by wave functions is useful for the
understanding of many interesting quantum phenomena, such as tunneling \cite{hartman1962}, 
interferences \cite{perez2001}, resonances \cite{susan1994}, electron transport \cite{zhou2010}, etc.,
and presents some similarity with other wave phenomena such as the transmission of light in 
materials \cite{garcia2010}. 

In this work, we take the most tractable case, the plain square barrier, that is an standard 
set-up for many theoretical purposes \cite{cohen} and can be useful for our present goal,
namely, to check the behaviour of different statistical magnitudes in the scattering process 
of quantum particles.   

The calculation of information theory measures in quantum bound states has been performed for different  
systems in the last years \cite{panos2005,sanudo2008-,ferez2009,lopezruiz2011,sanudo2012}.
These statistical quantifiers have revealed a connection with physical measures, 
such as the ionization potential and the static dipole polarizability \cite{sen2007} in
atomic physics. Other relevant properties concerning the bound states of 
atoms and nuclei have been put in evidence when computing these indicators on these many-body systems. 
For instance, the extremal values of these measures on the closure of shells \cite{panos2009,sanudo2009} 
and the trace of magic numbers \cite{lopezruiz2010,sanudo2011} are some of these properties.

The evaluation of these magnitudes in a quantum system requires the knowledge of the probability density 
as the basic ingredient. For bound states, this is directly known in some cases such as the H-atom \cite{sanudo2008-} 
or numerically derived in other cases from a Hartree-Fock scheme \cite{panos2007,borgoo2007}.
For no bound states, we proceed in this work to show how to perform this calculation.
We address this objective in the particular case of the scattering process of quantum particles through 
a potential barrier. The simplest obstacles which can be studied are the square barrier and 
the square well, two physical set-ups that can receive an equivalent mathematical treatment. 
In our case, the phenomenon of reflection (or transmission) of a wave function 
through a rectangular potential barrier in a one-dimensional set-up is considered \cite{cohen}. 
This standard system  presents three different regions depending on the value of the potential $V(x)$,

\begin{equation}
V(x) = \left\{
\begin{array}{rcl}
0, & \mbox{   }x\leq 0 & \mbox{(Region I),} \\
V_0, & \mbox{   }0<x<L & \mbox{(Region II),} \\
0, & \mbox{   }x\geq L  & \mbox{(Region III),} 
\end{array} \right.
\label{eq1}
\end{equation}

with $L$ and $V_0$ the width and height of the barrier, respectively.

When the free particle of mass $m$ encounters the barrier from the left for an energy $E>V_0$,
the solution $\phi(x)$ of the time-independent Schr\"odinger equation for the potential (\ref{eq1})
can be written as

\begin{equation}
\phi(x) = \left\{
\begin{array}{ll}
\phi_I(x)= & A_1\e^{ik_1x}+A_1'e^{-ik_1x}, \\
\phi_{II}(x)= & A_2\e^{ik_2x}+A_2'e^{-ik_2x},\\
\phi_{III}(x)= & A_3\e^{ik_1x}, \\
\end{array} \right.
\label{eq2}
\end{equation}

where there is no reflected wave ($e^{-k_1x}$ term) in the Region III. The expressions for
the wave numbers are: $k_1=\sqrt{2mE/\hbar^2}$ and $k_2=\sqrt{2m(E-V_0)/\hbar^2}$, with 
$\hbar$ the Planck's constant. Observe that when the particle comes in through the barrier 
with an energy $0\leq E\leq V_0$, the wave number $k_2$ becomes imaginary, 
then $\phi_{II}(x)= A_2\e^{\rho_2 x}+A_2'e^{-\rho_2 x}$, with 
$\rho_2=\sqrt{2m(V_0-E)/\hbar^2}$. The five amplitudes ($A_1, A_2, A_3, A'_1, A'_2$) are complex 
numbers determined, up to a global phase factor, by the normalization condition and the boundary 
constraints, namely the continuity of the wave function and its derivative at $x=0$ and $x=L$. 

The scattering region (Region II) provokes a partial reflection of the incident wave. The reflection
coefficient $R$ gives account of the proportion of the incoming flux that is reflected by the barrier.
The expression for $R$ is:
\begin{equation}
R = {Flux_{reflected} \over Flux_{incident}}={|A_1'|^2 \over |A_1|^2}\,.
\end{equation}
In this process, there are no sources or sinks of flux, then the transmission coefficient $T$ is given 
by $T=1-R$. 

It is straightforward to see that depending on the energy of the incident particles
there are two different behaviours in the scattering process, 
let us say the cases $0\leq E\leq V_0$ and $E>V_0$. 
If we write the energy of the particles as $E=pV_0$ with $p$ a non-dimensional parameter, 
then the cases to study are: $0\leq p\leq 1$ and $p>1$.

The reflection coefficient for $p>1$ yields:
\begin{equation}
R = {(k_1^2-k_2^2)^2\sin^2(k_2L) \over 4k_1^2k_2^2+(k_1^2-k_2^2)^2\sin^2(k_2L)}=
{\sin^2(k_2L)\over 4p\,(p-1)+\sin^2(k_2L)}\,,
\label{eq3}
\end{equation}
and for $0\leq p\leq 1$ is:
\begin{equation}
R = {(k_1^2+\rho_2^2)^2\sinh^2(\rho_2L) \over 4k_1^2\rho_2^2+(k_1^2+\rho_2^2)^2\sinh^2(\rho_2L)}=
{\sinh^2(\rho_2L)\over 4p\,(1-p)+\sinh^2(\rho_2L)}\,.
\end{equation}

\begin{figure}[t]
\centerline{\includegraphics[width=7cm]{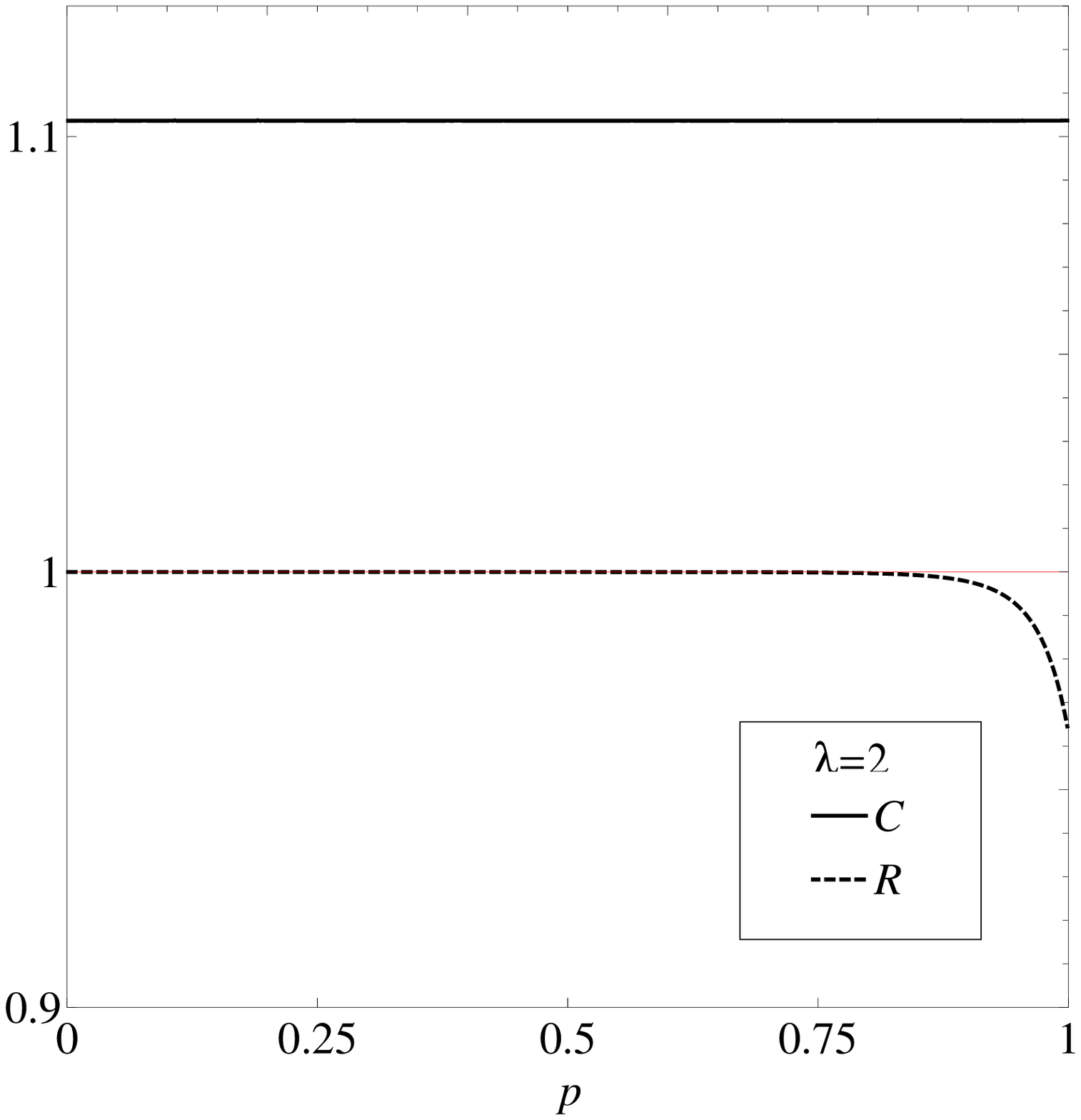}\hskip 5mm\includegraphics[width=7cm]{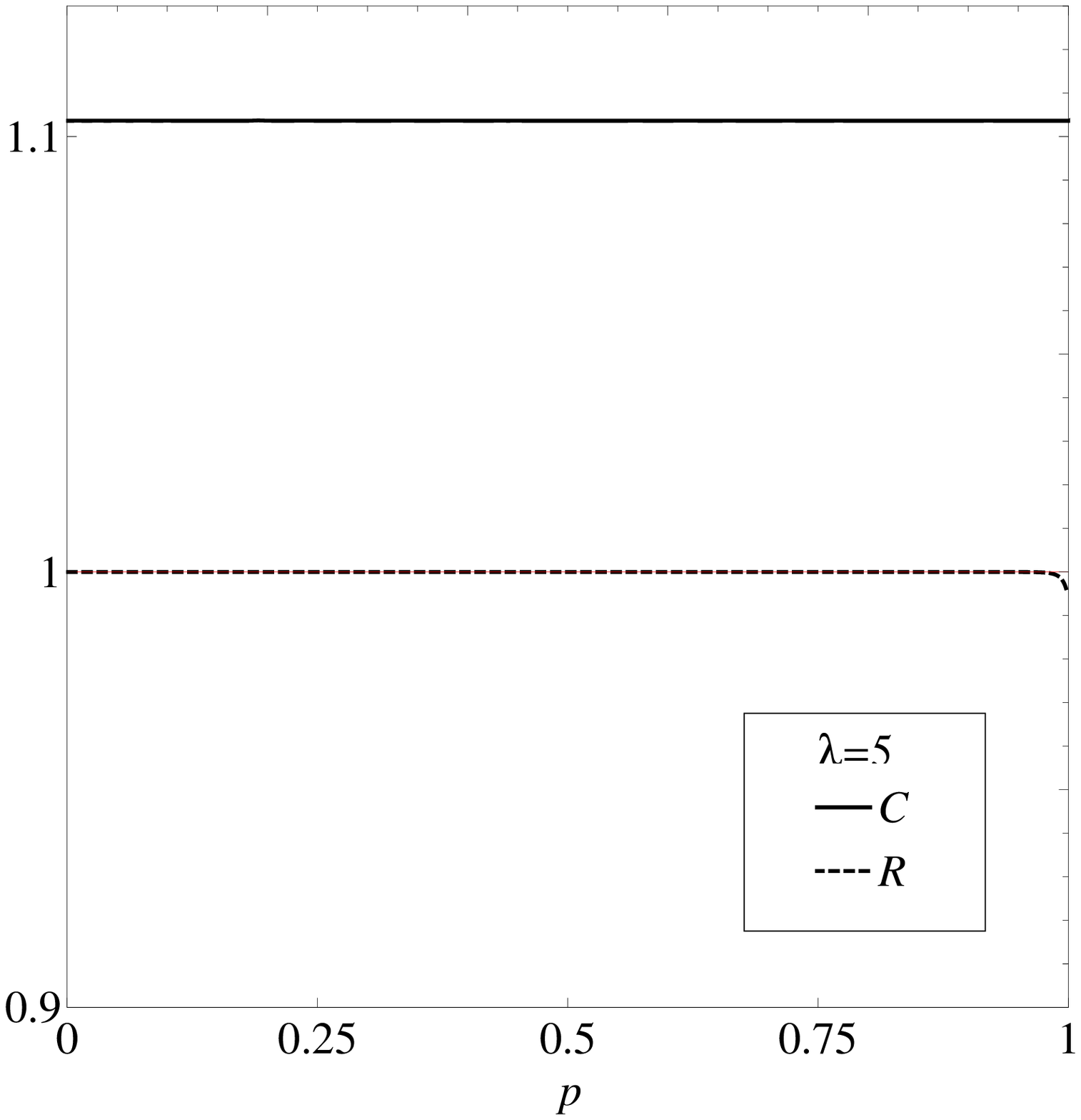}}
\centerline{(a)\hskip 7cm (b)} 
\caption{Statistical complexity, $C$, and reflection coefficient, $R$, vs. 
the dimensionless energy parameter, $p$, of the incident particle in Region I for $p<1$. 
(a) $\lambda=2$ and (b) $\lambda=5$. (The red line indicates the level value $1$).}
\label{fig1}
\end{figure}

\begin{figure}[t]
\centerline{\includegraphics[width=7cm]{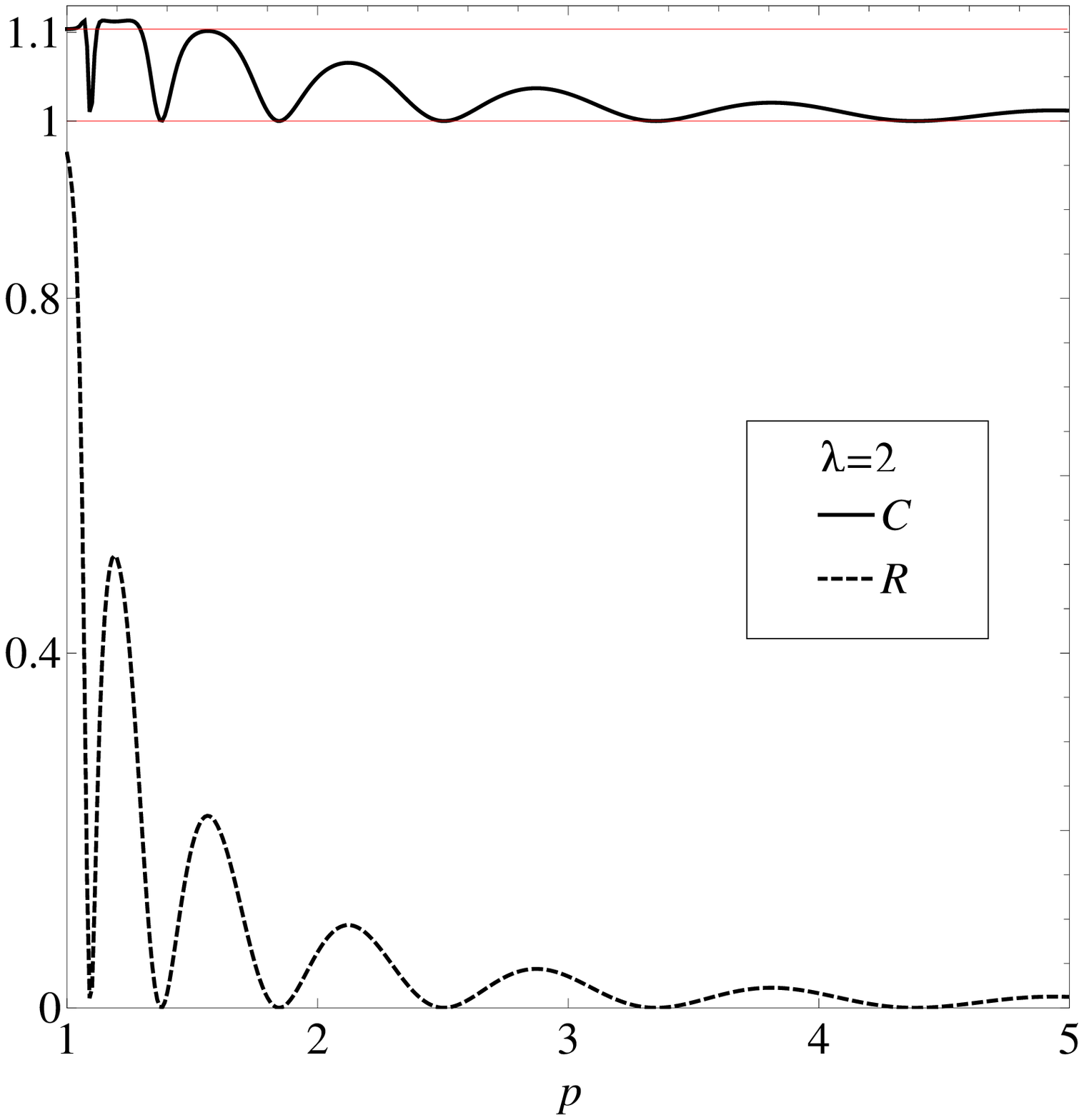}\hskip 5mm\includegraphics[width=7cm]{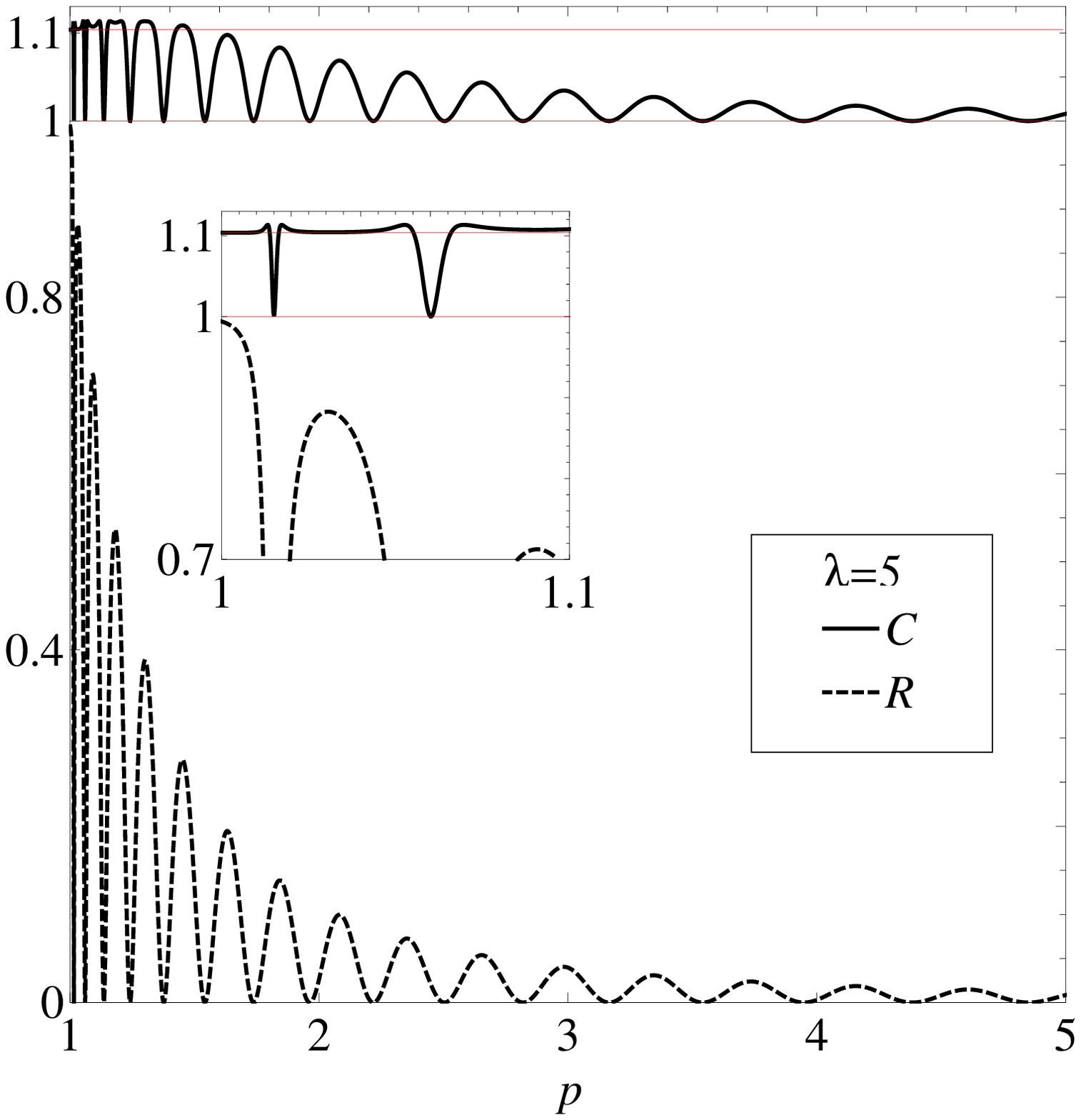}}
\centerline{(a)\hskip 7cm (b)} 
\caption{Statistical complexity, $C$, and reflection coefficient, $R$, vs. 
the dimensionless energy parameter, $p$, of the incident particle in Region I for $p>1$. 
(a) $\lambda=2$ and (b) $\lambda=5$. Detail for $p\gtrsim 1$ in the inset.
(The red lines indicate the level values $1$ and $3/e\simeq 1.10$).}
\label{fig2}
\end{figure}

\begin{figure}[t]
\centerline{\includegraphics[width=7cm]{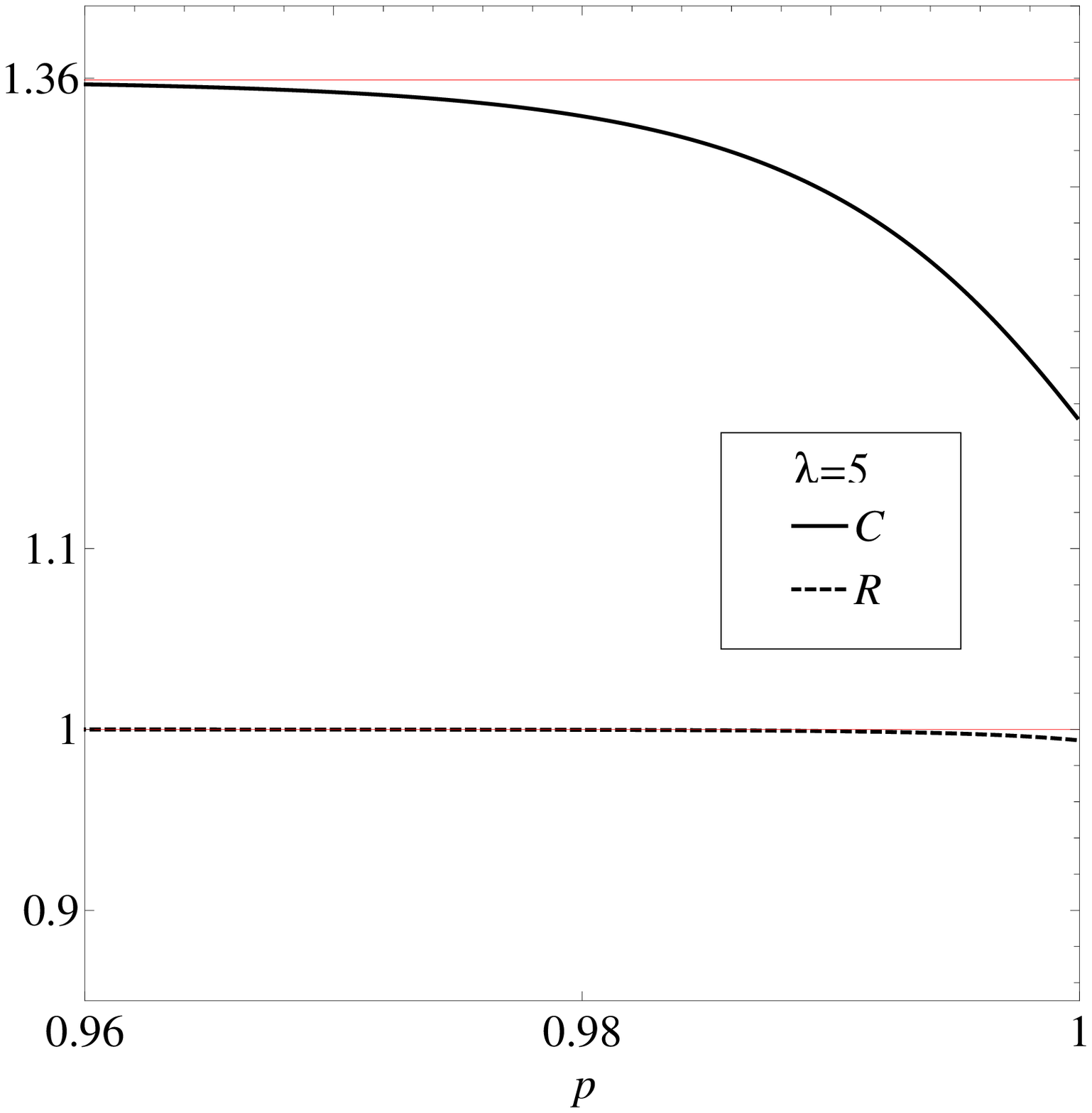}\hskip 5mm\includegraphics[width=7cm]{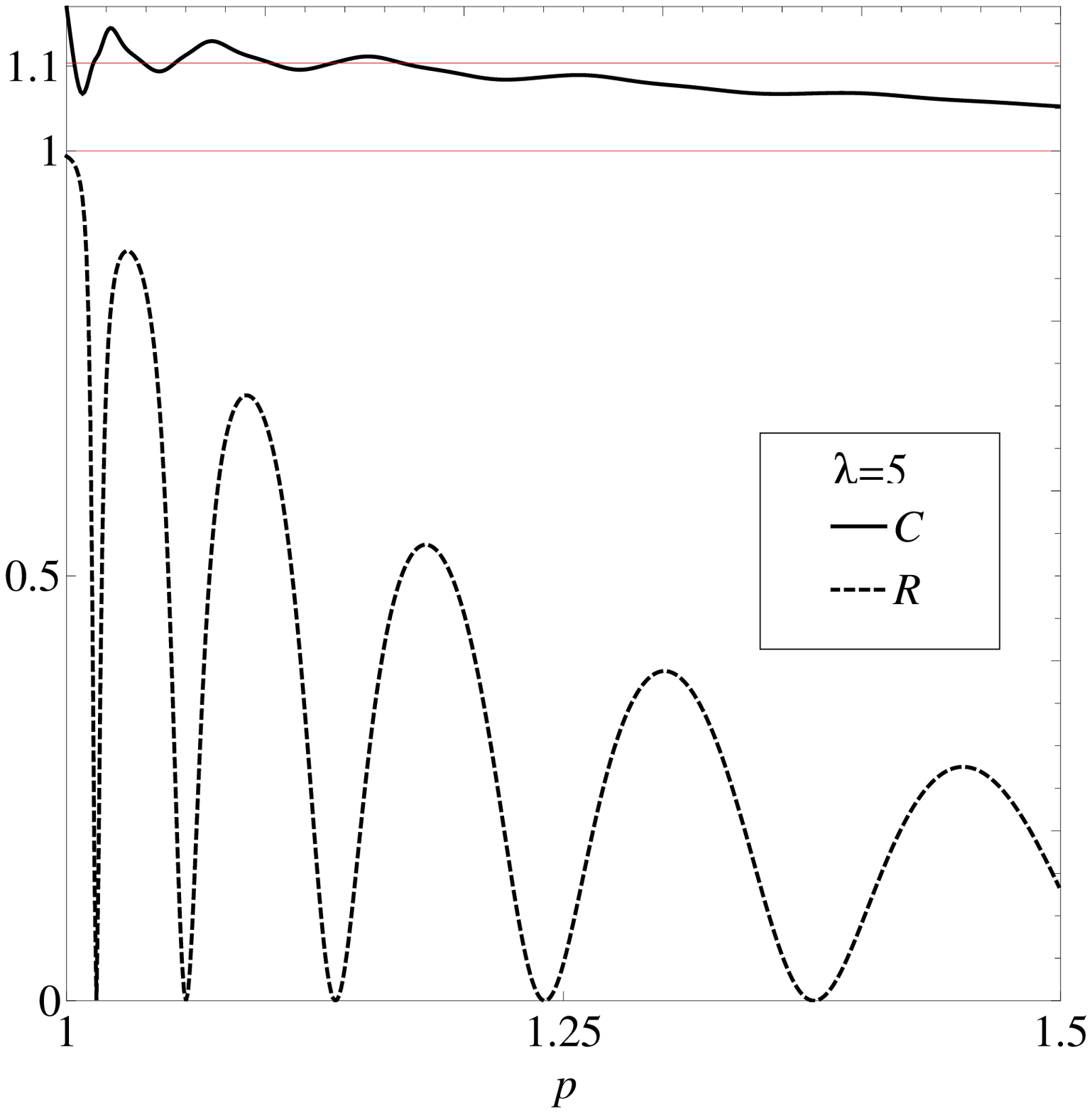}}
\centerline{(a)\hskip 7cm (b)} 
\caption{Statistical complexity, $C$, and reflection coefficient, $R$, vs. 
the dimensionless energy parameter, $p$, of the incident particle in Region II 
for (a) $p<1$  and (b) $p>1$, with $\lambda=5$. 
(The red lines indicate the level values $1$, $3/e\simeq 1.10$, $e/2\simeq 1.36$).}
\label{fig3}
\end{figure}

\begin{figure}[t]
\centerline{\includegraphics[width=8cm]{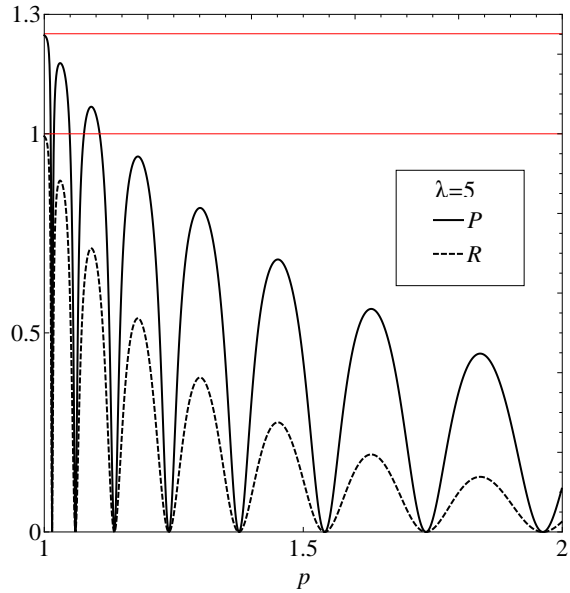}}     
\caption{Fisher-Shannon information, $P$, and reflection coefficient, $R$, vs. 
the dimensionless energy parameter, $p$, of the incident particle 
in Region I for $p>1$ with $\lambda=5$. 
(The red lines indicate the level values $1$ and $1.252$).}  
\label{fig4}  
\end{figure}

In order to compute the reflection coefficient $R$, it is necessary to give some concrete values 
to the size of the barrier and the mass of the particles. For the plots presented in Figures 
\ref{fig1}-\ref{fig4}, we have taken  $V_0=1$ eV, $L=\lambda\,L_0$ with $L_0=10$ \AA\; and $\lambda$
a positive real constant, and $m=0.511$ MeV the electron mass. For these values, we find that
\begin{equation}
k_2L = 5.123\,\lambda\,\sqrt{p-1}\,,
\end{equation}
and
\begin{equation}
\rho_2L = 5.123\,\lambda\,\sqrt{1-p}\,.
\end{equation}

We also proceed to calculate two statistical magnitudes for this problem, the statistical complexity
and the Fisher-Shannon entropy. These magnitudes are the result of a global calculation done on
the probability density $\sigma(x)$ given by $\sigma(x)=|\phi(x)|^2$, taking into account that
the interval of integration must be adequate to impose the normalization condition in the wave function. 
Particularly, this interval of integration is taken to be $[-a,0]$, $[0,L]$ and $[L,L+a]$, with $a=\pi/k_1$,
for Regions I, II and III, respectively.

The statistical complexity $C$ \cite{lopez1995,lopez2002},
the so-called $LMC$ complexity, is defined as
\begin{equation}
C = H\cdot D\;,
\end{equation}
where $H$ is a function of the Shannon entropy of the system and $D$ gives account of the sharpness
of its spatial configuration. Here, $H$ is calculated according to the simple exponential 
Shannon entropy $S$ \cite{lopez2002,shannon1948,dembo1991}, that has the form, 
\begin{equation}
H = e^{S}\;,
\end{equation}
with
\begin{equation}
S = -\int \sigma(x)\;\log \sigma(x)\; dx \;.
\label{eq1-S}
\end{equation}
For the disequilibrium $D$, we take some kind of distance to the equiprobability distribution 
\cite{lopez1995,lopez2002}, that is,
\begin{equation}
D = \int \sigma^2(x)\; dx\;.
\label{eq2-D} 
\end{equation}

The Fisher-Shannon information $P$ \cite{romera2004,sen2008} is defined as   
\begin{equation}
P= J\cdot I\,,
\end{equation}
where the first factor is a version of the exponential Shannon entropy \cite{dembo1991}, 
\begin{equation}
J = {1\over 2\pi e}\;e^{2S}\;,
\end{equation}
with the constant $2$ in the exponential selected to have a non-dimensional $P$.
The second factor
\begin{equation}
I = \int {[d\sigma(x)/dx]^2\over \sigma(x)}\; dx\;,
\end{equation}
is the so-called Fisher information measure \cite{fisher1925}, that quantifies the roughness 
of the probability density.

The reflection coefficient, $R$, and the statistical complexity, $C$, for the low energetic 
particles, $0<p<1$, in Region I are plotted in Fig. \ref{fig1}. For small $p$, there is no penetration
of the flow and the particles are reflected in the barrier, that is, $R=1$. The interference between the incident 
and the reflected waves generates standing waves in this Region I, given that both of them have the same wave number
and the same amplitudes. The complexity of any standing wave is $C=3/e\simeq 1.1036$. This value also corresponds to
the complexity calculated for the eigenstates of the infinite square well \cite{lopezruiz2009}. 
When the energy of the particles approaches the height of the barrier, i.e. $p\lesssim 1$, the tunnel effect
becomes perceptible and some transmission through the barrier takes place, then $R\lesssim1$. 
It can be clearly seen in Fig. \ref{fig1}a better than in Fig. \ref{fig1}b due to the different widths 
of the barrier, $\lambda=2$ and $\lambda=5$, respectively. Despite the tunnel effect, the most of the flow
is reflected in the barrier and the standing waves are maintained in the Region I, then $C$ does not register
any change.

In Fig. \ref{fig2}, the behaviour of $R$ and $C$ for particles with higher energies than 
the height of the barrier, i.e. $p>1$, is shown in Region I. First, the continuity of $R$ and $C$ 
for $p=1$ is observed with respect to the values taken in Fig. \ref{fig1}. Second, the transmission 
of particles becomes more important as their energy increases. In the limit $p\gg 1$, the totality
of the flow goes through the barrier, then there is no reflected wave and $R$ decays to zero with
a power law, $p^{-2}$, as it can be obtained from Eq. (\ref{eq3}). Third, the quantum nature of the 
problem appears in the oscillatory behaviour of the reflection coefficient. When the condition of 
standing wave in the barrier is reached, that is, $k_2L=n\pi$ with $n=1,2,\ldots$, the barrier becomes
transparent and the totality of the flow is transmitted, then $R=0$. The values of the energy that
fulfil this condition are given by the following series of $p$ values:
\begin{equation}
p=1+\left({\pi \over 5.123\,\lambda}\right)^2\,n^2,  \;\;\;\; n=1,2,3\ldots
\label{eq-p}
\end{equation}
Observe that the density of zeros for $R$ increases with $\lambda$, the width of the barrier,
as it can be seen in Figs. \ref{fig2}a and \ref{fig2}b, where $\lambda=2$ and $\lambda=5$, respectively.
Finally, observe that $C$ also presents an oscillatory behaviour with an asymptotic decay to $C=1$ 
when $p\gg 1$. Remark that $C$ takes its minimum value $C=1$ just on the transparency points $p$, 
given by the series of values (\ref{eq-p}), where the particles, similarly to the case $p\gg 1$,
are plane waves in the Region I and then they generate a constant density on this region, which is 
the situation of minimum complexity. In between the transparency points, for $p\gtrsim 1$, 
the particles are still reflected in some proportion able to reproduce standing-like waves in the Region I,
and then $C$ takes a value near $C=3/e$, which is the complexity of the eigenstates of the 
infinite square well. (See the inset of Fig. \ref{fig2}b).

In Fig. \ref{fig3}, the behaviour of $R$ and $C$ is plotted for the Region II. 
Evidently, $R$ is a global property of the system and it takes the same values than
in the Figs. \ref{fig1} and \ref{fig2}. For low energies, $p<1$, the particles 
penetrate the barrier with an exponential decay. The complexity of the exponential
distribution is $C=e/2\simeq 1.36$ \cite{lopez2002}, which is the value taken by $C$ 
for $p\rightarrow 0$ (Fig. \ref{fig3}a).
When $p$ increases, the particles can pass through the barrier and the density changes 
from the exponential distribution toward the possibility, for $p\gtrsim 1$, 
of standing waves in the Region II (Fig. \ref{fig3}b).
Here, $C$ oscillates around the value $3/e\simeq 1.10$, which corresponds to the value of 
transparency points where $R=0$ and all the particles are transmitted.
For $p\gg 1$, the particles become planes waves everywhere and the complexity tends 
to the value $C=1$.

In the Region III, the system behaves as a plane wave travelling to the right for all 
the energies. Then, the behaviour of $C$ is trivial: $C=1$.

In Fig. \ref{fig4}, the behaviour of the Fisher-Shannon information $P$ is plotted 
for the Region I in the range $p>1$. The value of $P$ for $p<1$ tends to the
value of the Fisher-Shannon information of the standing waves, $P=1.252$,
that are present in Region I for the low energy case. For $p>1$, the oscillatory
behaviour explained for $C$ is also found for $P$, that follows the same pattern
of behaviour of the reflection coefficient $R$. In fact, $P$ takes a null value on
their minima that are also located on the transparency points (\ref{eq-p}).
When $p\gg 1$, $P$ decays to the value that corresponds to the Fisher-Shannon 
information of the plane waves, that is, $P=0$.

In summary, the behaviour of statistical indicators in a scattering process of quantum 
particles has been studied. The relationship of these indicators with a physical 
magnitude, the reflection coefficient, has been disclosed. We have put in evidence 
that the situations where the transmission through the barrier is complete, 
i.e. the transparency points (\ref{eq-p}), can be detected by measurements 
of the statistical magnitudes, in the same way as it can be done by measurements 
of the incident and reflected flux of particles, i.e. by the reflection coefficient.
The reason is that all these magnitudes show their minima on the transparency points 
in the free region (Region I), just the region where the measurements should be performed. 
Take into account that the discreteness of the transparency points series is the more 
evident consequence of the quantum effects.
Therefore, the investigation of these entropic measures for this scattering process 
has revealed again that certain quantum properties can also be detected by means
of these indicators.

\section*{Acknowledgements}
 The authors acknowledge some financial support from  the Spanish project
 DGICYT-FIS2009-13364-C02-01. J.S. also thanks to the Consejer\'ia de 
 Econom\'ia, Comercio e Innovaci\'on of the Junta de Extremadura (Spain) for 
 financial support, Project Ref. GRU09011.

\end{document}